\documentstyle[twoside,fleqn,espcrc2,epsf]{article}

\newcommand{\AmS}{{\protect\the\textfont2
  A\kern-.1667em\lower.5ex\hbox{M}\kern-.125emS}}
\newbox\grsign \setbox\grsign=\hbox{$>$} \newdimen\grdimen \grdimen=\ht\grsign
\newbox\simlessbox \newbox\simgreatbox
\setbox\simgreatbox=\hbox{\raise.5ex\hbox{$>$}\llap
     {\lower.5ex\hbox{$\sim$}}}\ht1=\grdimen\dp1=0pt
\setbox\simlessbox=\hbox{\raise.5ex\hbox{$<$}\llap
     {\lower.5ex\hbox{$\sim$}}}\ht2=\grdimen\dp2=0pt
\def\simge{\mathrel{\copy\simgreatbox}}
\def\simle{\mathrel{\copy\simlessbox}}

\title{ Topological Defects in an Open Universe}

\author{D.N. Spergel and U. Pen
	\address{Princeton University Observatory, Princeton NJ 08544}%
        \thanks{Research supported by NSF AST88-58145 (PYI),
	AST90-20863, NASA NAGW-2448 (ATP) and the NSF Grand Challenge
	Cosmology Consortium}
       }

\begin{document}

\begin{abstract}

This talk will explore the evolution of topological defects
in an open universe.  The rapid expansion of the universe
in an open model slows defects and suppresses the generation
of CBR fluctuations at large angular scale as does the altered
relationship between angle and length in an open universe.
Defect models, when normalized to COBE in an open
universe, predict a galaxy power spectrum consistent
with the galaxy power spectrum inferred from the galaxy
surveys and do not require an extreme bias.  Neither defect models
in a flat universe nor standard inflationary models
can fit either the multipole spectrum or the
power spectrum inferred from galaxy surveys.
\end{abstract}

\maketitle

\section{ORIGIN OF DENSITY FLUCTUATIONS}

The big bang theory, despite its successes in explaining
the Hubble expansion, the thermal cosmic background
radiation (CBR) and light element nucleosynthesis, is incomplete.
For the big bang theory, the notion that the geometry and evolution
of the universe is
described by the
Friedman-Robertson-Walker (FRW) metric, does not explain the origin
of the density fluctuations that grew to form today's galaxies
and large scale structure.

There are two classes of extensions of the standard big bang
model:  either the universe began
the FRW phase with density variations on scales that
were then superhorizon or the universe
began the FRW phase without fluctuations and subsequent
causal physics generated density fluctuations.  Models
with primordial potential fluctuations can be classified
as ``curvature models" as they begin the FRW phase
with variations in the space curvature.  The generation
of fluctuations in an initially smooth or ``isocurvature" universe
requires variations in the equation of state.
Inflationary
theories fall into the former category as they produce
curvature
fluctuations  during a deSitter phase
that predated the FRW phase.
Models with topological defects fall into the latter
category as strings or textures generate density fluctuations
in initially smooth universes.

There are a variety of physical mechanism that
can lead to spatial variations in the equation of state.
Peebles proposed that there are spatial variations in
the ratio of baryons to photons.  Since baryons
and photons have different equations of state, these
entropy fluctuations produce density fluctuations \cite{Peebles87}.
If there was a large-scale magnetic field, then its
dynamics would lead to large-scale variations in the
equation of state which would be another process
that could produce density fluctuations.  This
talk explores  topological defects, yet another mechanism that
generates density fluctuations in an initial smooth universe
through equation of state fluctuations.

\section{TOPOLOGICAL DEFECTS}

Topological defects can arise at a symmetry breaking in
the early universe.  Symmetry breaking is a fundamental
part of modern particle physics and most of us suspect
that the standard model of particle physics, with broken $SU(3)\times
SU(2)\times U(1)$ symmetry, is much simpler at higher energies.
This suggests that the universe underwent a series
of phase transitions and different regions
of space may be in different degenerate vacuum states.
As Kibble\cite{Kibble} noted,
different causal disconnected regions of space will
settle into different vacuum states.  These field misalignments
can generate topological defects
and can  generate density fluctuations  that
could have seeded galaxies\cite{Press,Vilenkin}

The type of defect that forms at a phase transition
depends on the homotopy group of the vacuum
manifold after the symmetry breaking.  This
can be seen by considering a simple potential
for a N-dimensional scalar field, $\vec \phi$:
\begin{equation}
V(\phi)= \lambda (\phi^2 - \phi_0^2)^2
\end{equation}
If $\phi$ is a one dimensional field, then different
regions of space will settle into one of the two
vacuum state, $\phi = \phi_0$ and $\phi = -\phi_0$.
Domain Walls will seperate these two vacuum states.
If $\phi$ is a two dimensional field, then the
vacuum manifold is $\phi = \phi_0 \exp(i\theta)$,
where  the phase angle $\theta$ varies over space.
Strings will form along lines at which $\oint \nabla\theta d\vec s
\ne 0$.  Monopoles, textures and non-topological
textures are associated with higher dimensional
vacuum manifolds (see table 1).

The cosmological behavior of a defect will depend
upon whether the symmetry is gauged or global.
While gauged cosmic strings and global strings
have very similar cosmological behavior, this
is not true of the other defects.  Gauged monopoles
are cosmologically dangerous--- they interact
so weakly that once formed, they do not annihilate
and can easily dominate the energy density
of the universe.  Global monopoles, on the other
hand, have significant long range interaction.
They  annihilate quickly enough so that their
energy
density is a constant fraction of the closure density
and are potentially interesting source of density
fluctuations\cite{BennettRhie,Barriol}.
Gauged textures do not
produce significant density flucutations, while
global texture are similar to monopoles
and are a potentially interesting source
of density fluctuations\cite{Turok89}.

\begin{table*}[hbt]
\setlength{\tabcolsep}{1.5pc}
\newlength{\digitwidth} \settowidth{\digitwidth}{\rm 0}
\catcode`?=\active \def?{\kern\digitwidth}
\caption{Types of Defects}
\label{tab:effluents}
\begin{tabular*}{\textwidth}{@{}l@{\extracolsep{\fill}}rrrr}
{Defect} & {N } & Homotopy Group& Global & Gauged \\
\hline

Domain Walls & 1 & $\Pi_0$ & x & \\
Strings & 2 & $\Pi_1$ & x & x \\
Monopoles & 3 &  $\Pi_2$ &x &    \\
Textures & 4 &  $\Pi_3$ & x &  \\
N.T. Textures & $N> 5$ & $\Pi_4, \Pi_5, ...$&  x& \\

\end{tabular*}
\end{table*}

\subsection{Evolution of Defects}

Strings, global monopoles and textures
in a flat universe have a relatively simple
dynamical evolution.  As the universe
expands, the defect network untangles
and the characteristic scale over which
the field changes significantly,
the coherence scale of the field, grows
linearly with conformal time.
Numerical simulations  of cosmic strings find
that the curvature scale of the strings is
a constant fraction of the
horizon size\cite{BennettBouchet,AlbrechtTurok,AllenShellard}.
In numerical simulations of global field evolution
in an expanding universe,
the characteristic scale of textures and the spacing
between textures is also found to grow linearly
with conformal time in a radiation and matter-dominated
universe\cite{Spergel90,Nagasawa92}
This scaling form has been used in many studies
of these models\cite{Gooding91,Albrecht92,Perivoropolous93,Spergel93}
to estimate the density fluctuations produced by these
defects.
This scaling form implies that the energy density
in the defects is a constant fraction of the closure
density.  If the scale of symmetry breaking
is $\phi_0$, then the energy density in
defects is $\delta \rho \sim \dot\phi^2 \sim (\phi_0/\eta)^2$, where $\eta$
is the age of the universe measured in conformal time, and
all our units are $\hbar=c=1$.
Since the closure density is $\rho \sim (M_{Pl}/\eta)^2$, this
implies that the energy density in defects is
scale invariant,
\begin{equation}{\delta \rho_{defect} \over \rho}
\simeq \left({\phi_0 \over M_{Planck}}\right)^2\end{equation}
Note that the energy density in defects is constant.
As the amplitude of density fluctuations produced
by defects is of order their energy density, the constant
energy density fraction in global field theories leads
to a scale-invariant spectrum of density fluctuations\cite{Press}, similar
to that predicted by inflation.  The amplitude of CBR fluctuations
detected by COBE\cite{Wright} and the amplitude of galaxy fluctuations imply
that $\delta \rho/\rho \sim 10^{-4} - 10^{-5}$.  Thus, if defects
are the source of density fluctuations, then $\phi_0 \sim 10^{-2} M_{Planck}$
suggesting a GUT scale phase transition.

\subsection{Spectrum of CBR Fluctuations}

On the large-angular scales probed by COBE, the dominant source
of microwave background fluctuations in most models are metric
fluctuations\cite{SachsWolfe}.  These fluctuations
can be divided into three classes, scalar fluctuations,
vector fluctuations and tensor fluctuations\cite{Bardeen,KodamaSasaki}.
The growing
and decaying matter modes  are scalar fluctuations.  The
vector fluctuations, vorticity fluctuations, are subdominant
in defect models and not predicted in inflationary models.
The tensor modes are the gravity wave fluctuations that
have attracted significant attention in the past
year in the context of inflationary models.

In defect models, the dominant source of microwave
fluctuations are scalar fluctuations.  Thus, we
can get a rough physical understanding for the
predictions of these models by focusing only
on fluctuations in the gauge-invariant Bardeen scalar
potential $\Phi$.  On scales smaller than the
horizon size, this potential is the familar
Newtonian gravitational potential.  Fluctuations
in the gravitational potential contribute to
the CBR fluctuations through two terms\cite{SachsWolfe,Gouda}:
\begin{equation}
{\delta T \over T}(\eta_0)
 \simeq  {\Phi \over 3}|_{\eta_{LS}}^{\eta_0}
+2\int_{\eta_{LS}}^{\eta_0} \frac{\partial\Phi}{\partial t} |_{x=x(t)} dt
+{\delta T\over T}(\eta_{LS})
\end{equation}
The first
term describes fluctuations at the surface of last scatter.
The second term describes microwave fluctuations produced as the photons
travel from the surface of last scatter to the observer.
$\eta_{LS}$ is the conformal time at the surface of last scatter
and $\eta_0$ is the conformal time today.
The third term describes contributions to the CBR due
to variations in the photon/baryon or photon/dark matter
ratio. In inflationary
models in a flat universe, only the first term is non-zero, thus, the CBR
fluctuations are probe potential fluctuations at the surface
of last scatter.  This has led to the oversimplifying
statement that COBE is probing density fluctuations at $z = 1000$.
This need not be true.  In defect models and all models
in open or $\lambda$-dominated universes, $\dot \Phi$ is non-zero.
In most of these models, the quadrupole is produced primarily
by the decay of potential fluctutations at $z \sim 1/\Omega $.
In the PIB model, the third term is an additional source
of CBR fluctuations.

Since defects tend to generate potential fluctuations predominantly on
their coherence scale, they
predominantly produce microwave fluctuations on angular
scales close to that subtended by the horizon at a given redshift.  In
a flat universe, $\theta_H \simeq (1+z)^{-1/2}$ radians.  Thus,
textures collapsing at $z \sim 50$ are the dominant source of CBR
fluctuations on angular scales around 10$^o$.  Since defects produce a
scale invariant spectrum of potential fluctuations below the coherence
scale, they also produce a scale invariant spectrum of fluctuations in
the microwave sky.  Thus, like inflation, they predict that the
amplitude of the harmonic multipole, $c_l$, scales as $l^{-2}$ at
large $l$.  However, at scales larger than the coherence scale today,
microwave fluctuations in the defects models are suppressed and $c_l
\propto l^{-1}$ rather than $\propto l^{-2}$ at small $l$.  As we will
see when we consider open universe models, this will lead to a
dramatic suppression of the quadrupole and other low multipole
moments.

\subsection{Spectrum of Density Fluctuations}

Since defect evolution is a scale-invariant process, defects
produce a density fluctuation spectrum similar to
that predicted in the simplest inflationary models.  As noted
earlier, defects tend to produce fluctuations primarily
at their coherence scale at a given epoch.  Since
density fluctuations basically do not start growing
until the universe is matter dominated, the power spectrum
of density fluctuations is peaked at the wavenumber associated
with the coherence scale at equality.  On scales larger
than the coherence scale at equality
and smaller than the coherence scale today,
the spectrum scales as $\propto k$ as the
process of generating fluctuations is scale-invariant over
this region.  On scales larger than the  coherence scale
today, the density spectrum scales as $k^4$ as there
has not yet been time for any causal process to produce
density fluctuations on these large scales.  Defects
collapsing in the future will produce fluctuations on
these scales.

On scales smaller than the coherence
scale at matter-radiation equality,  the shape of the
density spectrum is determined by the nature of the
dark matter.  Thus, over the range of scales probed
by observations of large-scale structure, the shape
of the power spectrum in  a CDM-dominated defect model
is similar to that in a CDM-dominated inflationary model.
However, in the defect model, the peak of the power
spectrum occurs at a somewhat smaller scale as the
field coherence scale is smaller than the horizon
size in these models.  This reduction of large
scale power is particularly dramatic in a string-seeded
$\Omega =1 $ models\cite{Albrecht92}.  As we will see, this loss
of large-scale power is likely to be deadly for these models.

\subsection{Numerical Simulations of Defects in Flat Universe}

While the rough physical arguments of the previous
section yield a qualititative understanding
of the predictions of defects models,  quantitative
comparision with observations requires large-scale
numerical simulations.  While I will focus
on results from simulations described in detail in \cite{Pen94},
the basic approach is similar for all defect simulations
\cite{Press89,BennettRhie93,Albrecht92}.

The first part of the simulation is the evolution of the defect
in the expanding universe.  There are two
basic approaches to field evolution,
either the Lagrangian of the fundamental
theory is used to determine the field evolution equation
in the expanding universe\cite{Press89}
or the field is evolved as a free field
with the constraint that $|\vec \phi| = \phi_0$
\cite{Turok89,BennettRhie93,Pen94}.
Both approaches yield similar results.

Contrary to what one might naively expect, the complexity of the
numerical simulation actually decreases with increasing $N$.  For $N
\longrightarrow \infty$, the field evolution can be solved exactly
\cite{TurokSpergel},
and only has small departures from Gaussianity at $N > 6$\cite{Jaffe}.
At $N=4$ and below, real defects exist in space time, causing singular
field configurations and the associated numerical uncertainties.  For
textures, the problem only arises at a small number of events in space
time which we model with the ``spin flipping'' technique\cite{Pen94}, but for
global monopoles the singularity is real and needs to be smeared out
over a few grid cells.  For strings, the situation is much more
difficult, because global strings carry an energy density proportional
to $\log(r_c/r_\xi)$ the log of the core radius (a GUT scale) to the
mean string separation (a fraction of today's visible universe).  This
ratio is nearly constant, while in the simulation we do not have
nearly enough dynamic range.

The field evolution determines the stress energy tensor
of the defects, which serves as a source  for the generation
of metric fluctuations.  In the numerical simulations, all
three types of metric fluctuations (scalar, vector and tensor)
are computed.  By evolving the scalar fluctuations to today,
we can compute the predicted amplitude of density fluctuations
in a given model.  The other metric fluctuations terms are
needed to compute $\delta T/T$.

Photons are propogated along geodesics towards an observer
placed somewhere in the simulation.  As the photon moves
towards the observer, the metric fluctuations  produce
temperature fluctuations that are summed along the
photon path.  In recent simulations, Coulson  et al.\cite{Coulson93}
have also included the effects of photon scattering off of
free electrons,  this scattering damps fluctuations
on angular scales smaller than the horizon size
at the surface of last scatter.  At the end of a simulation,
the microwave sky map can be compared directly
to the COBE observations.  The COBE detection
of fluctuations on the $10^o$ scale is used to
normalize the thoeries and to compare the theory
predictions to observations of large-scale structure.

The multipole spectrum of CBR fluctuations found
in the numerical simulations is consistent with
our expectations for a curvature theory.
On scales smaller than $\sim 60^o$, the angular
scale corresponding to the coherence scale today,
the CBR fluctuations are scale invariant.  This
implies that $c_l l (l+1)$ is constant.  On the
largest angular scales, there is a weak suppression
of the dipole, quadrupole and octopole.  However,
these would be difficult to detect given
the cosmic variance.
The predicted spectral shape, similar to
that predicted by inflationary scenarios, is consistent with the
current observations.
While the fluctuations predicted for COBE
are mildly non-Gaussian, the current observations
can not distinguish this signature.  High
signal-to-noise measurements on
small angular scale will be needed to detect
this signature of topological defects\cite{Coulson93}.

\begin{figure}[htb]
\vspace{9pt}
\epsfxsize=\hsize \epsfbox{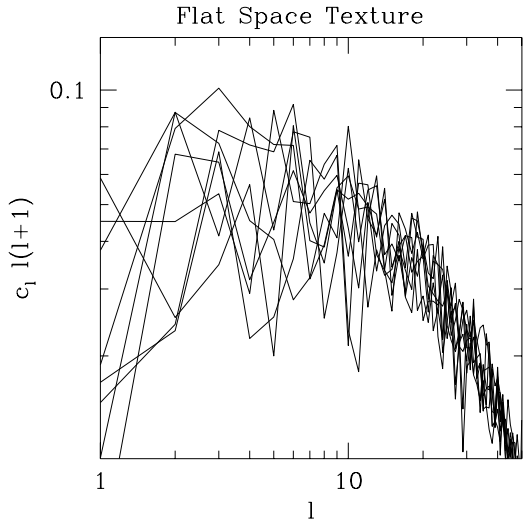}
\caption{Fluctuations from Global Texture in an $\Omega=1$ universe}
This figure from Pen et al. (1994) shows the calculated
CBR multipole spectra in a global texture simulation in
a flat universe.
The different lines correspond to the CBR
fluctuations seen by different observers at
various locations in the simulation box.  The
dispersion in their multipole measurements
in a signature of the expected cosmic variance.
\label{fig:CBR}
\end{figure}

\subsection{Comparison with Observations}

Unfortunately, the results for most defect
theories in a flat universe are quantitatively
disappointing.
These models in a flat CDM-dominated
universe normalized to COBE predict low amplitudes
of mass fluctuations on the 8/h Mpc scale
of $0.25 h_{50}^{-1}$\cite{Pen94}; these large bias factors
is somewhat concerning.  Even more concerning
for the theory is that it underpredicts the
amplitude of mass fluctuations on the 20/h Mpc
scale by a factor of 6! Albrecht and
Stebbins\cite{Albrecht92} reach similar conclusions
for cosmic string models in a CDM-dominated
universe.  This model also underpredicts
the amplitude of mass fluctuations at the 20/h Mpc
scale and appears unlikely to be consistent with
the APM\cite{Baugh}, IRAS\cite{Fisher93} and CfA\cite{Vogeley}
surveys.  Because of the
small coherence length of cosmic string theories,
their power spectra peak at smaller scales than
``standard" inflationary models, thus, we suspect
that more detailed analysis of the strings
+ HDM scenario will also conclude that
it does not  predict large enough fluctuations
at scale $> 20$/h Mpc.    All of these discrepancies, of course,
rely on numerical simulations.  It is certainly possible
that more detailed simulations will remove the discrepancies
between theory and observations.  They may also exacerbate
the discrepancies.

Defect theories are not alone in having difficulties fitting the
observed galaxy distribution.  Most inflationary scenarios can not
simulateously fit the observed level of COBE fluctuations and the
small pair-wise velocity dispersions observed at small scales.  This
has lead large-scale structure simulators into their baroque phase as
once simple theories acquire additional free parameters (e.g., mixed
dark matter, cosmological constants, tensor modes, ``designer"
spectra).

We believe that these discrepancies should encourage
us to reconsider our assumptions and explore open
universe models.  There are a host of observations
that point towards an open universe and as we will
see these models are remarkably successful
at explaining the observed large scale structure.

\section{OBSERVATIONAL EVIDENCE FOR LOW $\Omega$}
	Observations of the dynamics of clusters, their
gas content and the age problem all provide strong
motivation for considering open models.

	There are several independent techniques that can be used to determine
the mass-to-light ratios of clusters: measurements of the velocity dispersion
of galaxies, observations of X-ray temperature profiles and modelling of
the observed pattern of gravitational lensing of background objects by
a cluster. All of these techniques yield similar estimates for the mass
of clusters and imply that if mass traces light on the scales of clusters,
$\Omega$ is $\sim 0.2$\cite{Peebles93}
For example,
detailed studies of the dynamics of the Coma cluster finds
an M/L of $310 \pm 50 h$  on the scale of $5/h$ Mpc\cite{Hughes89}.
If this large region is a fair sample of the universe, then
$\Omega \simeq 0.2$.

	Measurements of the ratio of the mass of hot gas in clusters to
the total mass of a cluster imply that baryons comprise at least
1/6 of the mass of the cluster for $H_0 = 50$ km/s/Mpc\cite{White91}.
Combining this with estimates of the baryon density from
from big bang nucleosynthesis\cite{Walker93}
$\Omega_b h^2 \simeq 0.015$, suggests that $\Omega << 1$.  These
observations provide further evidence that the universe is open
or that the baryons are not very good tracers of the mass on
the scale of clusters of galaxies.

	Observations of the galaxy motions also provide evidence that
$\Omega << 1$.  On scales of $\sim 1$Mpc, the measured pairwise
velocities of galaxies, $\sim 300$ km/s, implies
that $\Omega \simeq 0.15 \exp(\pm 0.4)$\cite{Davis83}.
Numerical simulations of CDM models normalized to
COBE  predict a pairwise dispersion of $1000 \Omega^{0.6}$ km/s.
The thinness of the structure in the observed large-scale structure
such as the ``Great Wall"\cite{Geller89}
the caustic structure seen in superclusters\cite{Regos89}
as well as the cold local Hubble
flow inferred from
observations of galaxy peculiar velocities\cite{Ostriker90},
provide additional evidence that the large small-scale random velocities
predicted in an unbiased $\Omega= 1$ model are not observed.

	Observations of galaxy motions on large scales do, however,
hint at larger values of $\Omega$.  While  analysis of the redshift structure
of the IRAS galaxy sample of galaxies imply that
$\Omega^{0.6}/b_{IRAS} \sim 0.5$\cite{Fisher94}
larger densities are suggested by the comparison of the galaxy
velocities with the distribution of IRAS galaxies using the POTENT
algorithm\cite{Dekel93}: $\Omega^{0.6}/b_{IRAS} \sim
1.28_{-0.59}^{+0.75}$, where $b_{IRAS}$ is the ratio of the
fluctuations in the IRAS selected galaxies to the ratio of the
fluctuations in the underlying mass distribution.  The POTENT estimate
is subject to uncertainities due to the non-linear Malmquist biases
\cite{Gould}
Since the IRAS sample is known to be biased against
elliptical galaxies, it is quite plausible that optical galaxies are
better tracers of the underlying mass distribution than infrared
galaxies.  Thus, if we use the observed ratio of optical to infrared
galaxy fluctuations\cite{Lahav90}, $b_{optical}/b_{IRAS} = 1.7$
even these observations suggest $\Omega < 1$ and are compatible with
$\Omega \sim 0.2$.

	Requiring that the age of the universe is older than  that of
stars in globular clusters provides additional evidence that $\Omega << 1$.
Most recent measurements of the Hubble constant suggest
that $H_0 \sim 75$ km/s/Mpc\cite{Jacoby92}.  On the other hand, the
age of the universe, $t_0$, should exceed 13 billion years,
the lower limit estimated for the age of globular
clusters\cite{Deli89}
These results suggest that $H_0 t_0 \ge 0.8$.  This is in
excess of the predictions of a flat ($\Omega =1)$
matter dominated universe, $H_0 t_0 = 2/3$, and is more
consistent with either an open universe, or a universe
whose energy density is dominated by a cosmological constant.

	These arguments suggest that there is strong astronomical motivation
for considering model in which $\Omega < 1$.  This need not imply
that the universe is open.  If the universe is dominated by a cosmological
constant, then the  universe could still be flat.  It is difficult, however,
to understand the origin of such a small cosmological constant,
$\Lambda \simeq 10^{-128} M_{Planck}^4$.
Thus, we will focus on open models with $\Lambda = 0$.

\section{DEFECTS IN AN OPEN UNIVERSE}

\subsection{Analytical Model}

	As a first step towards exploring the evolution and
effects of defects in an open universe model,
a simple analytical model provides a useful estimate the density fluctuations
produced by defects\cite{Spergel93}.  Comparision with numerical
simulations (see figure 2) suggest that this model may be
an accurate description of the density fluctuations produced
by defects.

\begin{figure}[htb]
\vspace{9pt}
\epsfxsize=\hsize \epsfbox{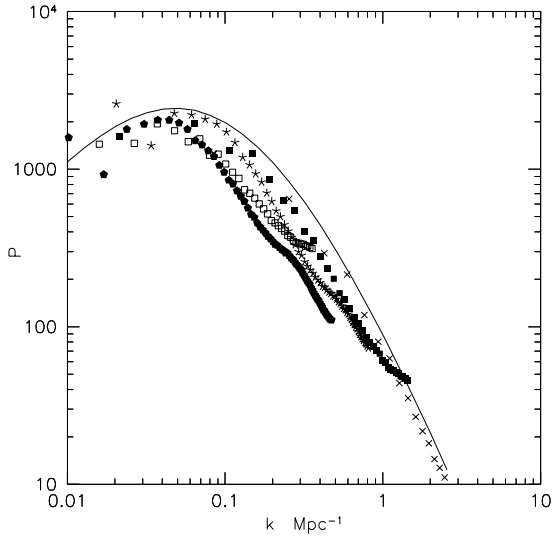}
\caption{Density Fluctuations (Analytical Model)}
This figure  from Spergel(1993) compares the
density fluctuation spectrum predicted by an analytical model
for the gravitational effects of defects with numerical
simulations (Pen et al. 1994).  The solid line is
the predictions of analytical theory and the triangles
and squares are the results of various simulations with
different grid sizes.
\label{fig:Density}
\end{figure}

	Since defects generate density fluctuations through
variations in the equation of state, the gravitational
effects of defects can be approximated by describing
the evolution of the pressure fluctuations produced by
defects, $\delta p$.  On scales larger than the defect
coherence scale, $l_{coh}$, the pressure fluctuations
produced by the defects are uncorrelated.  Thus, for
small $k$, $\delta p$ is independent of $k$.  On
scales smaller than the coherence scale, $\delta p$
is suppressed.  Thus, defect effects can be described by,
\begin{equation}\delta p(k) = \Theta(\beta k \eta - 1)\end{equation}
where $\beta$ is the ratio of the defect coherence scale to the
horizon size.  For texture models, $\beta \sim 1/3$, while for string
models, $\beta \sim 1/5 - 1/10$.  Once we have described the evolution
of pressure fluctuations, the gauge-invariant approach of Kodama and
Sasaki\cite{KodamaSasaki} can then be used to compute the evolution of
the gravitational potential and density fluctuations.  A similar
approach has been used by Albrecht and Stebbins\cite{Albrecht92} and
Perivoropolous\cite{Perivoropolous93} to
model the generation of density fluctuations by cosmic strings.

	This analytical model suggest that defects are extremely
promising.   The spectrum is similar to that reported
based on the reanalysis of the COBE data\cite{Wright93} and
the COBE normalized theory implies reasonable galaxy density fluctuations,
$\sigma_8 \sim 0.5 \pm 0.2$ for $\Omega = 0.2$ and $h = 0.8$.
This is consistent with determinations of $\sigma_8$
from the cluster mass function: $0.6 < \sigma_8 < 0.8$.
This success of the analytical model suggests that defects
in an open universe merit further investigation.

\subsection{Numerical Simulations of Defects in an Open Universe}

	On scales much smaller than the curvature scale, the same
numerical techniques that have been used to simulate flat universe
models can be easily rescaled to simulate open universe models.
However, on the large scales probed by COBE, significant changes must
be made in any algorithm in order to simulate the evolution of
potential fluctuations in negatively curved space.  We have developed
codes for simulating defects in the non-Euclidean negatively curved
open universe.

	We found that a convenient approach to evolving potential
fluctuations and microwave fluctuations in an open universe
was to use the Poincar\'e metric:
\begin{equation}
ds^2 = {dx^2 + dy^2 + dz^2 \over z^2} \qquad z > 0
\end{equation}
This metric has significant advantages over the more familiar
Robertson-Walker metric as it does not have a prefered position for
the observer and geodesics in this metric are simple analytical
expressions for all positions in space\cite{Wilson83}.  It is
straightforward to lay down evenly spaced grid points and it is
possible to make the grid periodic in x and y.  This semi-periodic
grid is the three dimensional analogue to the surface of
rotation generated by a tractrix.  In the z direction, we extend the
grid beyond the horizon of the observers and use mirror boundary
conditions.  While this is mildly wasteful of memory, it has the
advantage of simplicity.

	Texture, non-topological texture and monopole theories can be
accurately described on cosmological scales by a non-linear sigma
model\cite{Turok89}.  We use this model to evolve these theories using
the algorithms described in Pen et al.\cite{Pen94}.  At the beginning of
the simulations, when $\Omega \simeq 1$, the field in the open
universe simulations has approximately the same scaling energy density
as in the flat universe simulations.  As the universe begins to become
open and expand more quickly, the field oscillations are damped by
Hubble expansion and the scaling energy density drops. In a flat
universe, the scaling density implies $\dot{\phi}^2 \propto
(\nabla\phi)^2 \propto 1/\eta^2$.  When $\Omega = 0.2$, the kinetic
energy density has dropped to 70\% of its flat universe value.  This
lower energy density will lead to a suppression of CBR fluctuations on
large angular scales.  A decreased kinetic energy implies an increased
gradient energy, but since the integral is over $\dot{\Phi}$, it is
only the time dependent part which contributes to the Sachs-Wolfe effect.

	In our open universe simulations, we focus on the scalar
fluctuations produced by isotropic stresses.  In our flat
universe simulations, we found that these scalar fluctuations
were the dominant source of CBR fluctuations.  In the open
universe calculations, we will assume that the ratio
of tensor and vector fluctuations to scalar fluctuations is
the same as in the flat universe calculations and use
this to rescale our CBR fluctuations.  Since the textures
and monopoles move somewhat slower in an open universe,
this should suppress tensor and vector modes, thus, our
rescaled CBR fluctuations result may be slightly overestimated.
Throughout our open universe calculations, we use the gauge-invariant
formalism described in \cite{KodamaSasaki} to calculate
the potential fluctuations produced by the isotropic stresses.

	In our calculations, we compute these fluctuations by following
photons as they propogate towards an observer in the center of
the box and compute the Sachs-Wolfe contribution to
$\delta T/T$ (equation 3).  We then use a spherical
harmonic expansion to compute the amplitude
of each multipole moment.  The results in figure 3 are
for several different observers in the same box in
an $\Omega = 0.2$ texture-seeded universe.  The figure
shows the dramatic signature predicted by  open universe
models: a suppression of low multipoles.

\begin{figure}[htb]
\epsfxsize=\hsize \epsfbox{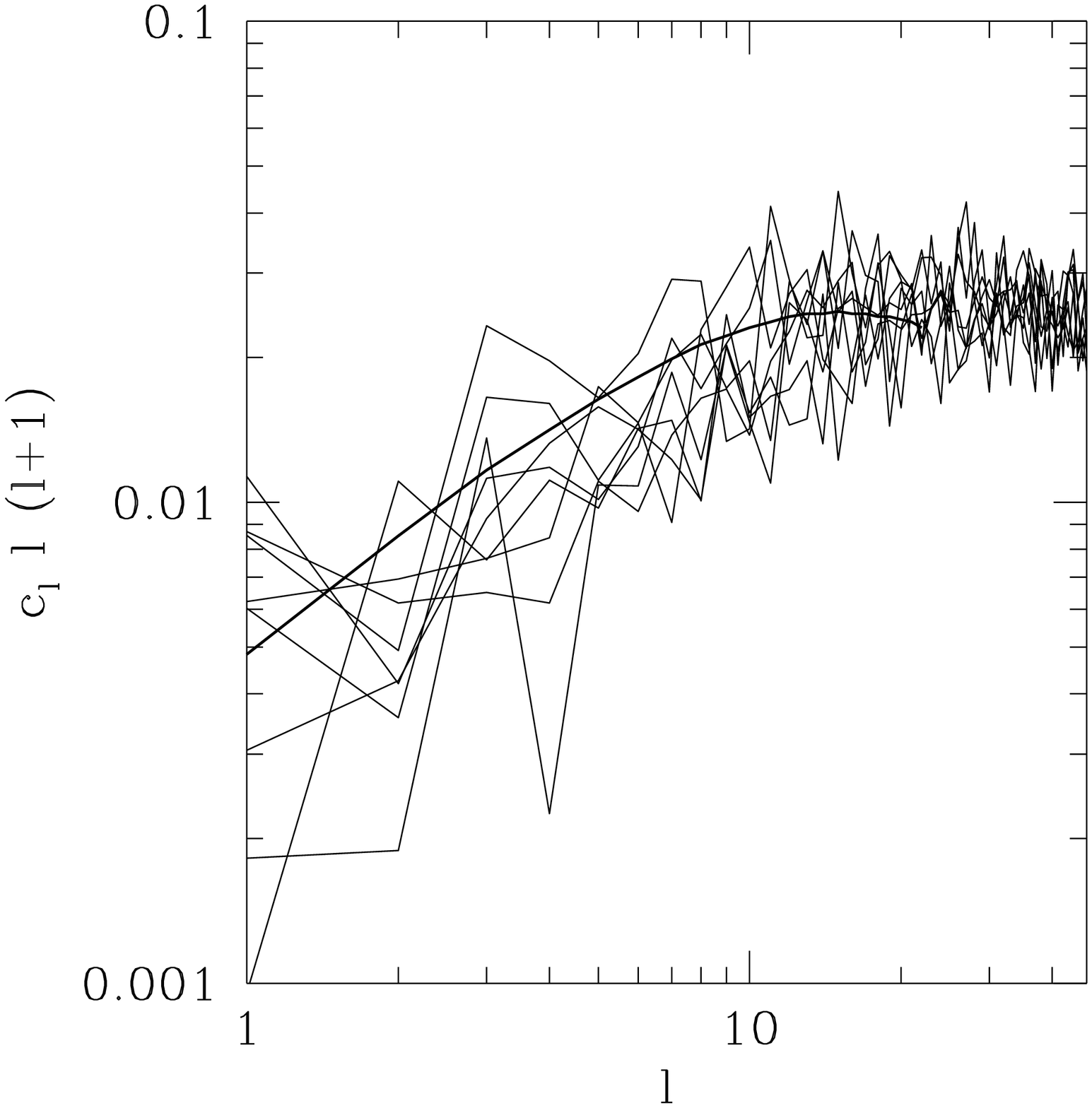}

\caption{}
The different solid curves trace the multipole
spectrum observed by different observers in a open
$\Omega = 0.2$ CDM dominated universe.  The temperature
fluctuations are normalized so that $8 \pi^2 G\phi_0^2 = 1$
with the tensor and vector multipole contributions scaled
from the flat universe model.  The heavy line
is the prediction of the analytic model\cite{Spergel93}.
\end{figure}

The relationship between angular scale and horizon
size at a given redshift,
$\theta \simeq \Omega (1+z)^{-1/2}$,  leads to
a characteristic signature for open universe models
with defects
on the large angular scales probed by COBE.  This effect
is more dramatic  than in open universe models that enter
the FRW phase with
curvature fluctuations\cite{Kamion}.
In a flat universe, regardless of whether fluctuations
are seeded by defects or by inflation, the multipole
fluctuations are expected to scale as
$c_l \propto 1/[l(l+1)]$ for $l \simle 50$.  Open
models follow this form for $l \simge 20$ as curvature was
relatively unimportant when fluctuations were generated
on these scales.  However, on large angular scales,
these models predict a suppression of CBR fluctuations.
This suppression of low multipoles contrasts with inflationary
models that predict an enhancement of low multipoles due
to gravitational waves\cite{Crittendon}.

The current observational results shown
are tantalizing but not yet definitive.  The COBE data\cite{Wright93}
suggest that the low multipole moments are suppressed,
a trend that appears to be confirmed by the larger
amplitude fluctuations seen on $l \sim 20$ by the MIT
experiment\cite{MIT} and $l \sim 30$ by the Tenerife experiment\cite{Ten}.
There remain large uncertainties in these measurements and further
observational work will be needed before
any strong statement can be made about the multipole spectrum.
It is, however, very intriguing that open universe models
make clear predictions at low $l$ and that the data
seems to follow this trend.

	Microwave predictions on smaller
angular scales depend on the ionization history
of the universe.  Without reionization, low $\Omega$ models
with scale-invariant spectra are not consistent with microwave
limits on small angular scales\cite{Bardeen87}.
If an early burst of star formation reionized the universe, then
microwave fluctuations are suppressed on
angular scales smaller than that corresponding to the horizon
scale at the redshift at which the universe was last optically
thick.  In a flat universe with $H_0 =50 $km/s/Mpc and $\Omega_b = 0.06$,
the surface of last scatter, $\tau = 2/3$ corresponds to $z = 60$
and microwave fluctuations are suppressed on scales smaller
than $\sim z_{LS}^{-1/2} \sim 6^o$, thus, these models
predict that fluctuations are suppressed for $l > 30$\cite{Crittendon}.
This early reionization appears to be inconsistent with several of the
reported detections of microwave fluctuations\cite{MAX,SK93,Python,MSAM}
on scales
of $\sim 1^o$.
In an
open $\Omega = 0.2$ universe with the same parameters, the surface
of last scatter corresponds to $z = 75$, and microwave fluctuations are
only suppressed below $\Omega_0 z_{LS}^{-1/2} \sim
1^o$.  This would be consistent with
both the detections at the $30' - 2^o$ scale and the limits
of $\delta T/T < 2.3 \times 10^{-5}$ recently obtained on angular
scales of 12'\cite{WD}.  As the observations
rapidly improve, we will hopefully be able to detect (or
rule out) the distinctive CBR signature of open universe
models.

\subsection{Density Fluctuations in an Open Universe}

The COBE observations can again be used to normalize
the one free parameter in defect theories for fixed
$\Omega$ and $h$: the scale of symmetry breaking, $\phi_0$.
With this parameter fixed, we can determine the predicted
galaxy power spectrum by scaling the results
of our earlier simulations of defects in a flat universe.
This calculation ignores the effects of baryons which will
slightly suppress the spectrum at scales smaller than
the Silk damping scale.

\begin{figure}[htb]
\epsfxsize=\hsize \epsfbox{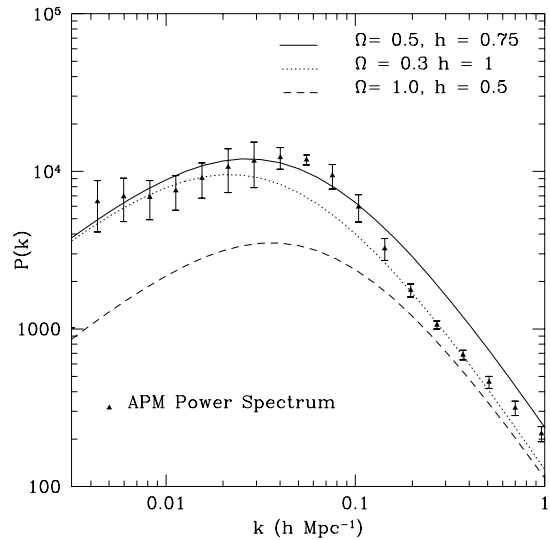}
\caption{Comparision of COBE normalized theories to APM
Galaxy Survey}
The open squares are the power spectrum of galaxy
fluctuations computed from the APM survey.
The lines show the predicted power spectrum for COBE-normalized
defect models with different values of $\Omega$ and $h$.
The bias is assumed to be 3 for the $\Omega =1$ and 2.5 for
the $\Omega = 0.3$ and $\Omega = 0.5$ models.  These bias
factors have a ``1 $\sigma$" uncertainty of  20\%.
\label{fig:COBE}
\end{figure}

The agreement between the texture-seeded open model and
the APM observations is remarkable.  The open model
appears to fit both the shape of the CBR spectrum
as determined from the COBE, MIT and Tenerife experiments
and the shape of the galaxy power spectrum.

While our numerical simulations of defects in an open
universe have focused on textures, the basic conclusions
are likely to be equally valid for strings, global monopoles
and non-topological texture.  Strings with their much
smaller coherence scale should predict an even larger
suppression of low multipoles and a ``break" in the
spectrum at large $l$ for fixed $\Omega$.
In a flat universe dominated by cold dark matter, the cosmic
string power spectrum is not a good fit to the observed
galaxy distribution\cite{Albrecht92},  however, the same
model rescaled to an open universe does remarkably well
at fitting the observations.  The required string tension,
$G\mu  \simeq 3 - 5\times 10^{-5}$ is roughly consistent with
the current estimate of string induced CBR fluctuations
\cite{Coulson93,Bennett92,Pen94}.   In an open universe,
the millisecond pulsar constraint on the energy density
in gravity waves implies a much weaker constraint on $G\mu$,
thus, these limits which are restrictive for $\Omega = 1$
are less troublesome for the model in an open universe.
We suggest that open universe string model deserves more
careful consideration.

Another intriguing model
is a low $\Omega$ baryons-only string-seeded model.
In this model,
string wakes would be the dominant mechanism for seeding
structure\cite{Vachaspati,Melott}.  We plan to investigate
this model in more detail.
Some further analyses and speculations are reported in
\cite{Spergel94}.

\section{CONCLUSIONS}

The generation of density fluctuations by defects
is a plausible alternative to the generation of
fluctuations by quantum fluctuations during inflation.
Defects in an open universe are a particularly exciting
model as texture in an open CDM dominated model appears to
be consistent with the observed CBR spectrum and the
observed level of galaxy fluctuations.

\end{document}